\documentclass{elsarticle}
\usepackage{times}
\usepackage{graphicx}
\usepackage{color}

\begin{document}

\title{Geometric phases in discrete dynamical systems}

\author[iact,ic1]{Julyan H. E. Cartwright}
\author[epfl]{Nicolas Piro}
\author[uib]{Oreste Piro}
\author[imedea]{Idan Tuval}

\address[iact]{Instituto Andaluz de Ciencias de la Tierra, CSIC--Universidad de Granada, E-18100 Armilla, Granada, Spain}
\address[ic1]{Instituto Carlos I de F\'{\i}sica Te\'orica y Computacional, Universidad de Granada, E-18071 Granada, Spain}
\address[epfl]{\'Ecole Polytechnique F\'ed\'erale de Lausanne (EPFL), 1015 Lausanne, Switzerland}
\address[uib]{Departamento de F\'isica, Universitat de les Illes Balears , E-07122 Palma de Mallorca, Spain}
\address[imedea]{Mediterranean Institute for Advanced Studies, CSIC--Universitat de les Illes Balears, E-07190 Mallorca, Spain}

\begin{abstract}
In order to study the behaviour of discrete dynamical systems under  adiabatic cyclic variations of their parameters,
we consider discrete versions of adiabatically-rotated rotators. Paralleling the studies in  continuous systems, we generalize the concept of geometric phase to discrete dynamics and
investigate its presence in these
rotators. For the rotated sine circle map, we demonstrate an analytical
relationship between the geometric phase and the rotation number of the system.
For the discrete version of the rotated rotator considered by Berry, the
rotated standard map, we further explore this connection as well as the role of
the geometric phase at the onset of chaos. Further into the chaotic
regime, we show that the geometric phase is also related to the diffusive
behaviour of the dynamical variables and the Lyapunov exponent.
\end{abstract}


\maketitle

Highlights:
\begin{itemize}
\item We extend the concept of geometric phase to maps.
\item For the rotated sine circle map, we demonstrate an analytical relationship between the geometric phase and the rotation number.
\item For the rotated standard map, we explore the role of the geometric phase at the onset of chaos.
\item We show that the geometric phase is related to the diffusive behaviour of the dynamical variables and the Lyapunov exponent.
\end{itemize}

\section{Introduction}

In continuous time dynamics, the study of adiabatic perturbations in general, and of adiabatic cyclic variations in particular,
is closely related to the concepts of anholonomy and geometric phase. The geometric phase \cite{shapere,chruscinski2004} is, indeed, a particular example of anholonomy that we can phrase as
the failure of certain variables to return to their original
values after a closed circuit in the parameters. Physical
expressions of such anholonomies appear in the rotation of the
plane of oscillation of a Foucault pendulum \cite{khein}, how
swimming is performed by microorganisms at low Reynolds numbers
\cite{shapere2},  how the stomach mixes \cite{arrieta2015},  and how a falling cat can manage to reorientate
itself in mid air in order to land on its feet \cite{montgomery1}.
The geometric phase was originally encountered --- as Berry's phase --- in
quantum mechanics \cite{berry3,berry2010}. From there, it was generalized to classical
integrable systems as Hannay's angle \cite{hannay}. Later it was extended to
nonintegrable perturbations of Hamiltonian systems \cite{montgomery,golin,spallicci}, and
thence to dissipative systems \cite{kepler2,kagan,tourigny2014}, all these instances within
the context of continuous-time dynamics. In the same context, rotated rotators
have been natural models in which to study this phenomenon because they provide
an easy way to control the adiabatic nature of the cyclic variation of the
parameter. With this in mind, Berry and Morgan, for instance, investigated the
geometric phase of a continuous-time Hamiltonian rotated rotator \cite{berry2}. In spite of the extensive research that has taken place in the last few decades
on geometric phases in a large class of applications, neither the geometric
phase nor any of its cognates have been considered hitherto in discrete
dynamical systems. Moreover, the general question of how a mapping-defined dynamics behaves under an adiabatic parametric cyclic perturbation has not been addressed until now. Our purpose in this paper is to make good this deficit and to introduce a discrete analogue of the geometric phase and show
that it is linked to important aspects of the dynamics of
maps. In order to do so, we adopt the paradigm of the rotated rotator but
follow an inverse sequence to the historical development. In the first place
we deal with the sine circle map that may be thought of as the discretization of a
kicked rotator of the type Berry and Morgan studied with the addition of strong
dissipation. We consider the results of discrete adiabatically-evolving parameter
loops in such a prototypical discrete-time dynamical system. The geometric phase in the
rotated circle map, it turns out, is intimately related to the behaviour of the
rotation number of the map as a function of the bare frequency parameter.
Turning to the Hamiltonian side, we study the rotated standard map, in which we
discover surprising relationships between the geometric phase, not only with the
rotation number as in the former case, but also with the Lyapunov exponent
and the diffusive behaviour of both action and phase variables. The reason for
our following this inverted developmental sequence is that 1D circle maps
are simpler as regards the transition from integrability to chaos, in comparison
with the much richer behaviour of the 2D Hamiltonian case.

\section{The rotated circle map}

\begin{figure*}[t]
\begin{center}
\includegraphics*[width=0.9\textwidth]{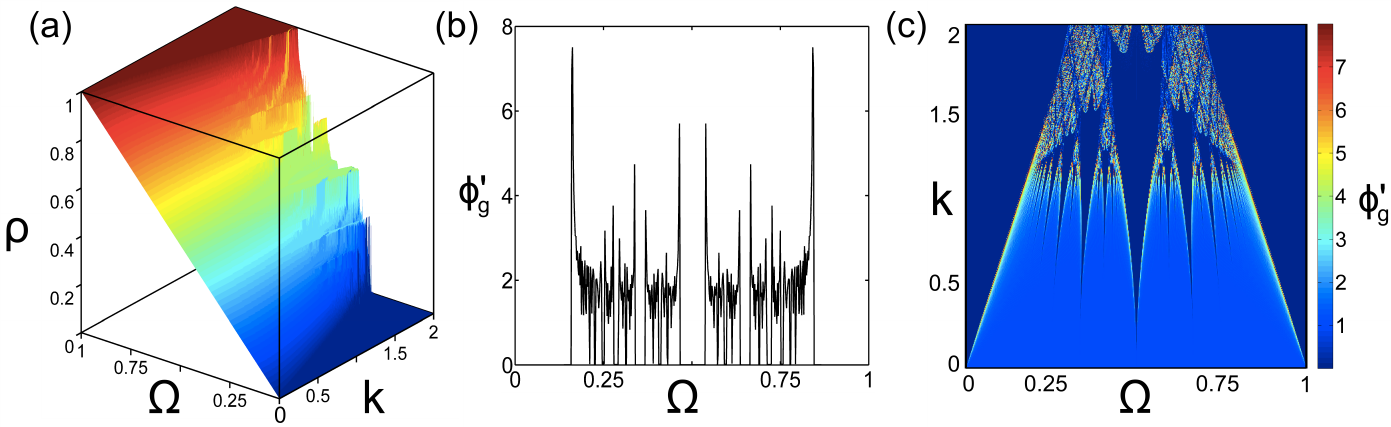}
\end{center}
\caption{ (a) Devil's quarry plot of rotation number $\rho$ in the
sine circle map; a section through the quarry with $k$ constant is
a devil's staircase. (b) Geometric phase $\phi_g$ plotted against
$\Omega$ for the rotated critical ($k=1$) sine circle map of
Eq.~(\ref{cv}). (c) In a colormap, same as in (b) but as a function of both parameters $\Omega$ and k.
}\label{devil}
\end{figure*}

Before introducing our rotated version of the circle map, let us first recall a
few necessary definitions and results on the original non-rotated one. The
circle map, usually written as
\begin{equation}\label{circlemap}
\theta_{n+1}=f^{n+1}_{\Omega,k}(\theta_0)=\theta_n+\Omega-\frac{k}{2\pi}\sin{2\pi\theta_n}\quad\bmod 1 ,
\end{equation} where $\theta_{n}=f^{n}_{\Omega,k}(\theta_0)$
represents the $n$th iterate of $\theta_0$,  which qualitatively
describes the dynamics of two interacting
nonlinear oscillators, is a one-dimensional
discrete mapping that describes how a rotator of natural frequency
$\Omega$ behaves when forced at frequency one through a coupling
of strength $k$. When $k=0$ the rotator runs uncoupled at
frequency $\Omega$, but when $k>0$ it can lock into a periodic
orbit: a resonance with some rational ratio $p/q$ to the driving
frequency. To measure the frequency of the rotator, i.e., the
average rotation per iteration of the map, it is useful to define
the rotation number
\begin{equation}
\rho=\lim_{N\to\infty}{\frac{\theta_N-\theta_0}{N}} .
\end{equation}
If we plot rotation number $\rho$ against $\Omega$ and $k$
--- a few hundred iterations after discarding an initial
transient are sufficient to give an accurate value for $\rho$
--- we obtain the devil's quarry \cite{bogpaper} illustrated in
Fig.~\ref{devil}(a). Periodic orbits with different rational
rotation numbers show up as so-called Arnold tongues: flat
steps in  Fig.~\ref{devil}(a). When $k<1$ the map is termed
subcritical, and intervals on which the rotation number is
constant and rational, where there is a periodic orbit of a
particular period, punctuate intervals of increasing rotation
number, whereas in supercritical circle maps ($k>1$) the
periodic orbits overlap. Chaos is found in the supercritical
circle map as iterates wander between the overlapping
resonances. In a critical circle map at $k=1$, at every value
of $\Omega$ there is a periodic orbit, and the rotation
number increases in a staircase fashion with steps at each
rational rotation number and risers in between. The devil's
quarry becomes the devil's staircase when we look at a
section with $k$ constant through the quarry. The ordering of
periodic orbits in the devil's staircase has been understood
in terms of Farey sequences and Stern--Brocot trees \cite{oreste,aronson,cvit,hao}, and
the transition to chaos in the circle map is well understood.

Now let us consider the rotated circle map; probably the simplest
discrete time system with a discretely and adiabatically varying
parameter. To this end we introduce a discrete slowly varying
parameter $X_n$:
\begin{eqnarray}\label{circlemap2}
&&\theta_{n+1}=\theta_n+\Omega-\frac{k}{2\pi}\sin{2\pi(\theta_n+X_n)}\quad\bmod 1, \nonumber \\
&&X_{n+1}=X_n+\Delta X=X_n\pm1/N
,\end{eqnarray}
where $n=1,2,\ldots N$, with $N\rightarrow\infty$ for adiabaticity.
Is there a geometric contribution to the phase after such an excursion?
Let us perform the change of variable $\theta'_n=\theta_n+X_n$, under which the map can be written as
\begin{eqnarray}\label{cv}
\theta'_{n+1}&=&\theta'_n+(\Omega+\Delta X)-\frac{k}{2\pi}\sin{2\pi\theta'_n}\quad\bmod 1\nonumber \\
&=&\theta'_n+\Omega'-\frac{k}{2\pi}\sin{2\pi\theta'_n}\quad\bmod 1
,\end{eqnarray}
where $\Omega'=\Omega\pm1/N$. So it is seen that the effect of the parameter loop is
just a shift in the value of $\Omega$.

In general, if one takes a
system through a parameter loop, one obtains as a result three
phases: a dynamic phase, a nonadiabatic phase, and a geometric phase.
If one then traverses the same loop in the opposite direction, the
dynamic phase accumulates as before, while the geometric phase is
reversed in sign. There is, of course, still the nonadiabatic phase
too; to get rid of this one must travel slowly around the loop.
Thence the geometric phase may be obtained as
\begin{equation}
\phi_g=\lim_{N\rightarrow \infty}{\frac{\phi_+-\phi_-}{2}}
,\end{equation}
where $\phi_+$ and $\phi_-$ are, respectively,  the total angles $\theta$ accumulated by travelling around the loop in the positive and negative directions. In terms of the primed variables, we can also define
\begin{equation}\label{hacv}
{\phi_g}' =\lim_{N\rightarrow
\infty}{\frac{f^{N}_{\Omega+1/N,k}(\theta_0)-f^{N}_{\Omega-1/N,k}(\theta_0)}{2}}
, \end{equation}
the obvious relation $\phi_g = {\phi_g}'-1$. Let us evaluate this limit, first for the simple
case $k=0$. Then $f^{N}_{\Omega,0}=\theta_0+N\Omega$, so
$f^{N}_{\Omega-1/N,0}=\theta_0+N\Omega-1$ and
$f^{N}_{\Omega+1/N,0}=\theta_0+N\Omega+1$, hence ${\phi_g}'=1$ and ${\phi_g}=0$.
This limiting case is conceptually equivalent to that of a Foucault
pendulum located at the Earth's equator where the plane of oscillations
remains fixed as the Earth rotates.

More interesting is what happens when $k\ne0$. From the definition of the rotation number
\begin{equation}\label{rn}
\rho=\lim_{N\rightarrow\infty}{\frac{f^{N}_{\Omega,k}(\theta_0)-\theta_0}{N}}
=\lim_{N\rightarrow\infty}{\rho^{N}_{\Omega,k}(\theta_0)}
,\end{equation}
where we are defining $\rho^{N}_{\Omega,k}(\theta_0)=(f^{N}_{\Omega,k}(\theta_0)-\theta_0)/N$, we have that
\begin{eqnarray}\label{gprn}
{\phi_g}'
&=&\lim_{N\rightarrow\infty}{\frac{f^{N}_{\Omega+1/N,k}(\theta_0)-f^{N}_{\Omega-1/N,k}(\theta_0)}{2}}\nonumber \\
&=&\lim_{N\rightarrow\infty}{\frac{\rho^{N}_{\Omega+1/N,k}(\theta_0)-\rho^{N}_{\Omega-1/N,k}(\theta_0)}{2/N}}
.\end{eqnarray}
We may recognize this as a peculiar type of derivative of the rotation number with respect to
$\Omega$ in which the limit $ N \to \infty $ of the definition of $\rho$ is taken simultaneously with the limit $\Delta\Omega \to 0$, maintaining $ N\Delta\Omega = 1 $. We may denote such a limit as $\bar\partial\rho/\bar\partial\Omega$. When both limits commute, this derivative coincides with the usual one and
\begin{equation}
{\phi_g}'=\frac{\bar\partial\rho}{\bar\partial\Omega}=\frac{\partial\rho}{\partial\Omega} ;
\end{equation}
so that we can say, in this sense, that the geometric phase for the
rotated circle map is the derivative of the rotation number. This is obviously
true within the tongues where the rotation number is constant and both
derivatives zero (Fig.~\ref{devil}(b and c)) and consequently ${\phi_g}'=0$ and
${\phi_g}=-1$. One might think of these cases as equivalents of a Foucault
pendulum located at one of the Earth's poles where the plane of oscillations
rotates one turn a day in the opposite direction (minus sign) to the Earth's
rotation. Outside the tongues both derivatives have singularities arising from
the peculiar dependence of $\rho$ on $\Omega$; nevertheless, apart from
convergence (or lack thereof) details, $\phi^\prime$ clearly has the
qualitative aspect of a derivative of the devil's staircase, as illustrated in
Fig.~\ref{devil}(b). Fig.~\ref{devil}(c), of which Fig.~\ref{devil}(b) is the
$k=1$ slice, shows also two more things: for $k<1$ where quasi-periodic
orbits are still measure-wise abundant the phase outside the tongues is rather
regular; this is a consequence of the well known ``trivial'' scaling relations of
the subcritical mode-locking structure \cite{shenker}. In the supercritical
region $k>1$ the limit is not well-behaved and fluctuations in the geometric
phase appear that correspond to the chaotic regions already noted in
Fig.~\ref{devil}(a).

\begin{figure*}[t]
\begin{center}
\includegraphics*[width=0.9\textwidth]{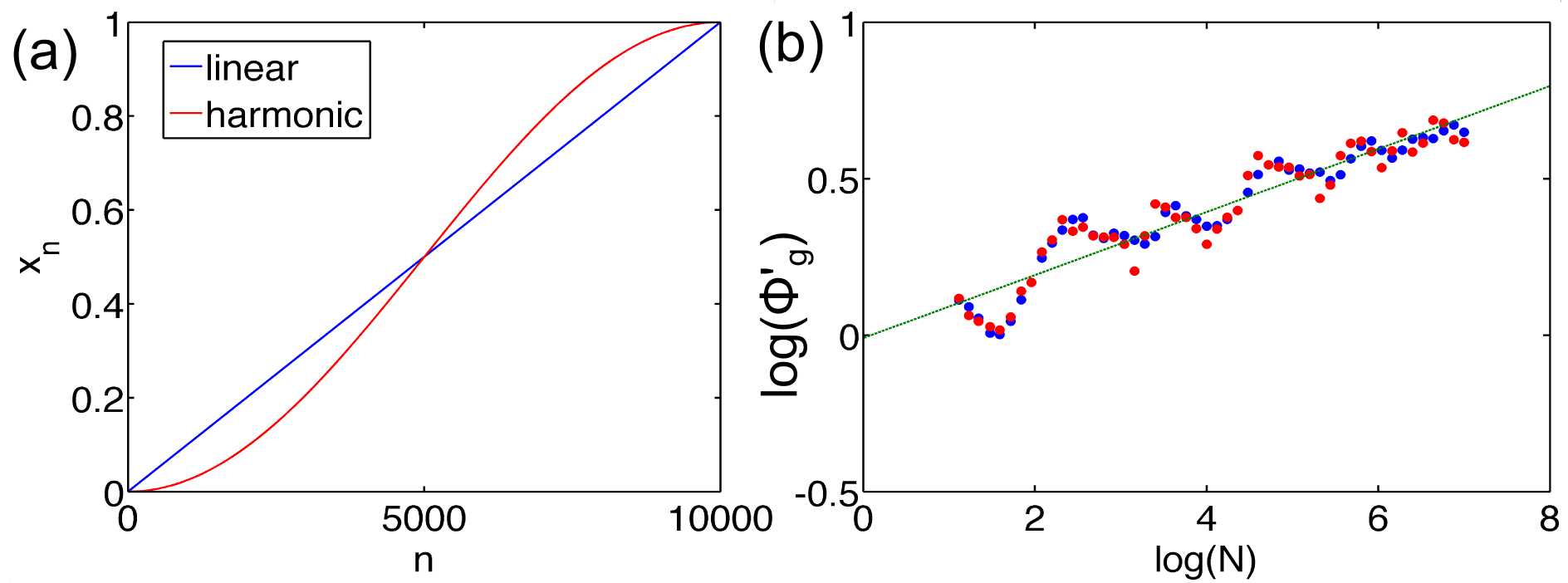}
\end{center}
\caption{(a) The two discretely and adiabatically varying parameter cycles used to illustrate the geometric nature of the phase. (b) In both cases, the geometric phase evaluated through eq.~(5) diverges in a similar manner for $\Omega=\pi/6$ (the line is a linear fit as a guide for the eye).}\label{harm}
\end{figure*}

We want to remark on an important difference between the behaviour of our discrete version of geometric phase and its continuous counterpart. While a characteristic feature of the latter is that it can assume any value between zero and $2\pi$, in the discrete case it can only take 0 or infinity. One might ask however, whether the places where the divergences occur depend or not on the shape of the adiabatic cycle in the parameter space. Therefore, to  assess in this way the geometric nature of the computed phases, we compare in Fig.~\ref{harm} the results obtained under two different protocols for the discrete slowly varying parameter $X_n$ over a cycle: the linear constant speed cycle considered above and a nonlinear sinusoidal cycle given by $X_n=(1-\cos(n\pi/N))/2$. Notice that in this case, the connection of our definition of the phase and the derivative of the rotation-number function is no longer analytically obvious. However, while there are clear differences between the corresponding values of the phases evaluated at finite adiabatic velocity variations or cycle lengths $N$'s  the positions on the $\Omega$  axis of both the null and diverging phases tend to coincide as the $N\rightarrow\infty$ limit is approached. A finer comparison of the divergent behaviour as a function of the shape of the cycle should be a subject of further studies.

\section{Berry's rotated rotator and the standard map}

Berry \cite{berry2} considered the geometric phase of the rotated
rotator $\ddot q= V_0\sin(q-X(t))$, that is
\begin{eqnarray}
&&\dot p= V_0\sin(q-X(t)), \nonumber \\
&&\dot q = p,
\end{eqnarray}
which is a rotator $\ddot q= V_0\sin q$ being rotated by $X(t)$.
In the same way that discretizing the rotator with mixed
forwards and backwards Euler methods gives us the area-preserving
Chirikov--Taylor standard map,
\begin{equation}
\mathbf{T}(I_n,\theta_n): \left\{\begin{array}{c}
I_{n+1} = I_n+(k/2\pi) \sin (2\pi\theta_n) \nonumber \\
\theta_{n+1} = \theta_n+I_{n+1}
\end{array}\right.
,\end{equation}
we can consider the rotated standard map that comes from the
rotated rotator
\begin{eqnarray}
&&I_{n+1}= I_n+(k/2\pi) \sin[2\pi (\theta_n+X_n)], \nonumber\\
&&\theta_{n+1} = \theta_n+I_{n+1}, ~ X_{n+1}=X_n\pm 1/N,
\end{eqnarray}
where $X_n$ is again a discretely and adiabatically varying
parameter that is moved around a closed loop from $0$ to $1$.

In the rotating frame of reference, where
\begin{equation}\label{prime_theta}
    {{\theta}'}_n={\theta}_n\pm X_n,
\end{equation}
the rotated standard map reads:
\begin{eqnarray}\label{rotated_sm_2}
  I_{n+1} =  I_n+(k/2\pi)\sin[2\pi(\theta_n\pm X_n)],\\ \nonumber
  {\theta}'_{n+1}\mp X_{n+1} = {\theta}'_{n}\mp X_{n}+I_{n+1},
\end{eqnarray}
or
\begin{equation}
{\theta}'_{n+1}= {\theta}'_{n} \pm (X_{n+1}-X_{n}) +I_{n+1}=
{\theta}'_{n}\pm\Delta+I_{n+1}.
\end{equation}
If we define $I'_n=I_n\pm\Delta$ we obtain the original standard map for the new primed variables:
\begin{equation}
   \left(\begin{array}{c}
   I'_{n+1}\\ \theta'_{n+1}
   \end{array}\right)=\left(\begin{array}{c}
   I'_n+(k/2\pi)\sin(2\pi\theta'_n)\\
   {\theta'}_{n}+I'_{n+1}
   \end{array}\right).
\end{equation}

\begin{figure}[tbp]
\begin{center}
\includegraphics*[clip=true,width=0.7\columnwidth]{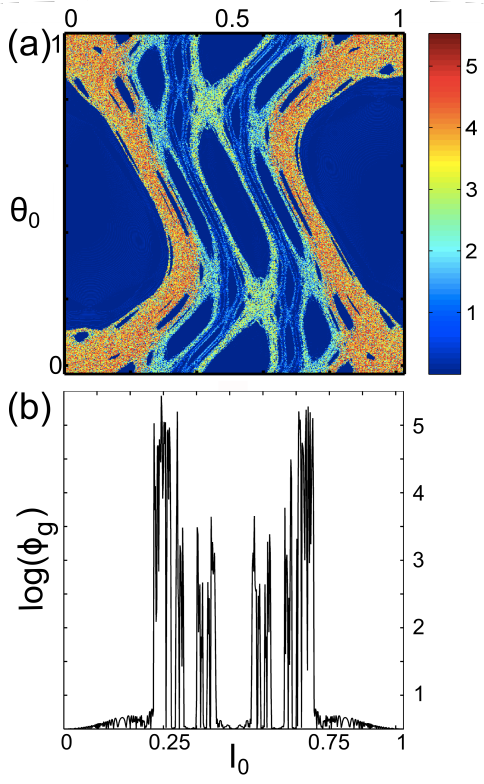}
\end{center}
\caption{\label{Mapastandard} The logarithm of the geometric phase for the rotated
standard map in a rainbow color code as a function of the initial
conditions ($I_{0},\theta_{0}$) for $k=1$. (b) A transversal cut through (a) at $\theta_{0}=0.5$ demonstrates that the phase is zero in the islands.}
\end{figure}

By the definition of the geometric phase,
\begin{equation}\label{StandMapGeomPhaseDef}
\phi_g=\lim_{N\rightarrow\infty}\frac{\left[{{\mathbf{T^+}}^N}\right]_\theta(I_0,\theta_0)
-\left[{{\mathbf{T^-}}^N}\right]_\theta(I_0,\theta_0)}{2}
\end{equation}
and
\begin{eqnarray}\label{StandMapGeomPhaseDef2}
&&\phi'_g= \\ \nonumber
&&\lim_{N\rightarrow\infty}\frac{\left[{\mathbf{T}}^N\right]_\theta(I_0+\Delta,\theta_0)-
\left[{\mathbf{T}}^N\right]_\theta(I_0-\Delta,\theta_0)}{2}=\\ \nonumber
&&\lim_{N\rightarrow\infty}\frac{\left[{\mathbf{T}}^N\right]_\theta(I_0+1/N,\theta_0)-
\left[{\mathbf{T}}^N\right]_\theta(I_0-1/N,\theta_0)}{2}.
\end{eqnarray}
Given that the rotation number $\rho$ can be expressed as
\begin{equation}\label{rot_number_sm}
    \rho(I_0,\theta_0)=\lim_{N\rightarrow\infty}\rho_N(I_0,\theta_0)=
    \lim_{N\rightarrow\infty} \frac{\left[{\mathbf{T}}^N\right]_\theta(I_0,
    \theta_0)-\theta_0}{N}
\end{equation}
we derive the following relation for the geometric phase:
\begin{eqnarray}
\phi'_g &=& \lim_{N\rightarrow\infty}\frac{\rho_N(I_0+1/N,\theta_0)-
\rho_N(I_0-1/N,\theta_0)}{2\left(1/N\right)} \nonumber \\
&=& \frac{\bar\partial{\rho(I_0,{\theta}_0})}{\bar\partial{I_0}}.
\end{eqnarray}

\begin{figure}[tp]
\begin{center}
\includegraphics*[clip=true,width=0.68\columnwidth]{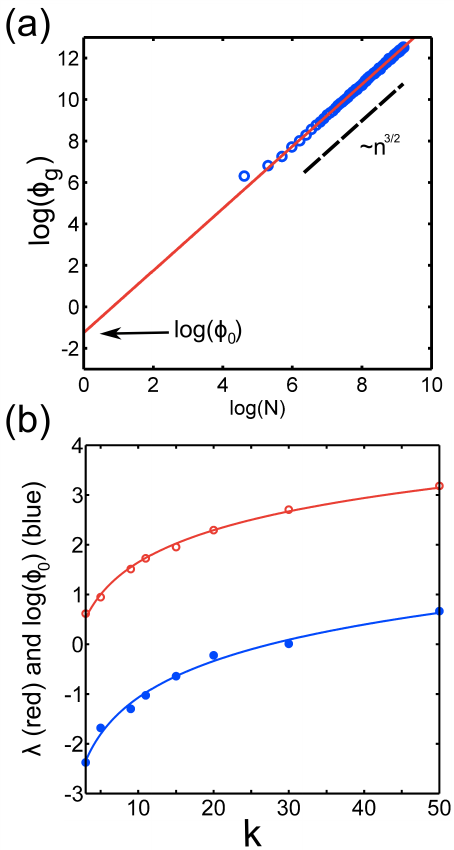}
\end{center}
\caption{\label{superdiffusion} The geometric phase for the rotated
standard map diverges superdiffusively for large $k$ with power-law
exponent $\beta\approx3/2$. (b) The $y$-intercept $\log(\phi_0)$
(filled circles) follows the logarithmic $k$-dependence of
the Lyapunov exponent, $\lambda$ (open circles) for large $k$.}
\end{figure}

Of course, this limit does not necessarily exist, nor can one expect it to be
finite, so we should take these equations as formal relations. However, as
Fig.~\ref{Mapastandard} shows,
within the regular KAM islands where the rotation number remains constant, this
relation is easily verified and the geometric phase is null. The situation is
much less intuitive within the chaotic regions of the phase space;
Fig.~\ref{Mapastandard} again. There, the
rotation number is not well defined and its dependence on the initial action
has not been studied. Naively one could estimate the difference between two
neighbouring trajectories in the numerator of the r.h.s of
Eq.~(\ref{StandMapGeomPhaseDef2}) as proportional to $2\exp(\lambda N)/N$, which
clearly diverges exponentially. On the basis of this assumption one could
attempt to establish a connection between this divergence of the geometric
phase and the Lyapunov exponent $\lambda$ of the trajectory as
$
\ln\phi'_g \approx N \lambda + \ln N
$.
However, the experimentally computed phase does indeed diverge with $N$, but at
a much slower pace than exponentially. Instead, we find power-law
behaviour for $\phi_g$ with an exponent of $3/2$, which is independent of the
value of the nonlinearity parameter $k$ (Fig.~\ref{superdiffusion}a). In other
words, the phase difference of neighbouring trajectories varying by an amount
$1/N$ in their initial actions grows superdiffusively at the same rate as that
of Richardson's law, after $N$ iterations. The failure of the exponential
assumption is because the Lyapunov exponent only holds if the double limit of
the initial perturbation going to zero and the time to infinity is taken in
this order. The phase, however, is defined with a similar double limit but
taken along the particular direction in which the product of time times the
initial perturbation is kept equal to one. On the other hand, the $3/2$ power law
can be explained from the well-known diffusive behaviour of the action due to
the chaotic dynamics of the standard map. From the equation for the phase we
may realize that the phase difference $\Delta\theta$ between two trajectories
evolves as
\begin{equation}\label{phase_difference_eq1}
    (\Delta\theta)_{n+1} - (\Delta\theta)_n = (\Delta I)_{n+1}.
\end{equation}
It follows, for large $n$, that
\begin{equation}\label{phase_difference_eq2}
    \frac{d|\Delta\theta|_{n}}{d n}  \simeq |\Delta I|_{n} \propto n^{1/2}
    \Rightarrow |\Delta\theta| \propto n^{3/2}.
\end{equation}
But there is a further surprising result in the behaviour of the phase
illustrated in Fig.~\ref{superdiffusion}b. By extrapolating the power law up to
$\ln N = 0$ we obtain the prefactor of the hypothesized $3/2$ power law; this
prefactor plotted as a function of the nonlinearity amplitude $k$ behaves very
similarly to the Lyapunov exponent plotted as a function of the same parameter,
up to an additive constant $\approx 2$. The extension, origin and consequences
of this connection are under current investigation. 

\section{Conclusion}

In summary, we have shown that the study of adiabatic modulations in maps and, in particular, the
extension of the concept of geometric phase to this type of systems opens up a
new avenue to understand their dynamical features.

\section*{Acknowledgements}

We acknowledge the financial support of the Spanish Ministerio de
Ciencia y Innovaci\'on grants FIS2013-48444-C2-1-P \& FIS2013-48444-C2-2-P; I.T. acknowledges a Ram\'on y Cajal fellowship.

\bibliography{geomphase}
\bibliographystyle{elsarticle-num}

\end{document}